\documentclass[aps,prd,reprint,preprintnumbers,superscriptaddress,showpacs,twocolumn]{revtex4-1}
\usepackage{latexsym,graphicx,amssymb,amsmath,mathrsfs}
\usepackage{setspace,bm}
\usepackage[breaklinks, colorlinks=true, pdfstartview=FitV, linkcolor=red, citecolor=blue, urlcolor=blue]{hyperref}
\usepackage[usenames]{color}
\usepackage{latexsym}
\usepackage{epstopdf}
\usepackage{mathtools}
\usepackage{amssymb}

\usepackage[utf8]{inputenc}

\usepackage{tikz}
\usetikzlibrary{shapes,arrows}

\allowdisplaybreaks[1]
\usepackage[normalem]{ulem}  

\renewcommand\sout{\bgroup \color{red} \ULdepth=-.5ex \ULset}
\begin{document}

\title{Information theoretical view of QCD effective model with heavy quarks}

\author{Kouji Kashiwa}
\email[]{kashiwa@fit.ac.jp}
\affiliation{Fukuoka Institute of Technology, Wajiro, Fukuoka 811-0295,
Japan}

\author{Hiroaki Kouno}
\email[]{kounoh@cc.saga-u.ac.jp}
\affiliation{Department of Physics, Saga University, Saga 840-8502,
Japan}

\begin{abstract}
To understand the phase transition phenomena, information theoretical approaches can pick up some important properties of the phenomena based on the probability distribution.
In this paper, we show information theoretical aspects of the 3-dimensional 3-state Potts model with the external field which is corresponding to the QCD effective model with heavy quarks.
The transfer mutual information which represents the information flow of two spin variables is numerically estimated based on the Markov-chain Monte-Carlo method.
The transfer mutual information has the peak near the confinement-deconfinement transition, and it may be used to detect the precursors of the transition. Since the transfer mutual information still have the peak even if the Polyakov-loop changes continuously and smoothly, we may pick up some aspects of the confinement-deconfinement nature from the information flow properties. Particularly, the transfer mutual information shows the significantly different behavior below and above the Roberge-Weiss endpoint existed in the pure imaginary chemical potential region, which may indicate the system change by the confinement-deconfinement transition.
\end{abstract}

\maketitle

\section{Introduction}

Understanding the confinement-deconfinement transition at finite temperature ($T$) and chemical potential ($\mu$) in quantum chromodynamics (QCD) is one of the important and interesting subjects in the elementary particle, nuclear, hadron and astrophysics.
In the ordinary understanding of the confinement-deconfinement transition at finite temperature in QCD with vanishing chemical potential, there is no ``phase transition'' and is the crossover. In the confinement-deconfinement crossover, there are no singularities in the local order-parameters and also the thermodynamic quantities. However, it has been recently discussed in Ref.~\cite{Sato:2007xc} that the confinement and deconfinement states at zero temperature can be clarified via the topological order~\cite{Wen:1989iv} and then it is not necessary that local order-parameters exist and several observable show singular behaviors for the confinement-deconfinement transition. The analogy of the topological order has been applied to the thermal QCD by employing the imaginary chemical potential, $\mu=(0,\mu_\mathrm{I})$, and then it is expected that confinement-deconfinement transition may be determined from the topological viewpoint~\cite{Kashiwa:2015tna,Kashiwa:2016vrl,Kashiwa:2017yvy}; see also Ref.\,\cite{Weiss:1987mp} for the pioneering work with the imaginary chemical potential and Ref.\,\cite{Kashiwa:2019ihm} for the review of the imaginary chemical potential. In the case of the chiral symmetry breaking which is another important nature of QCD, one interesting investigation from the information theoretical view was done by using the thermodynamic geometry~\cite{Castorina:2019jzw,Zhang:2019neb}.
These studies indicate that we need several viewpoints to correctly understand the confinement-deconfinement nature of QCD. 

In this study, we investigate the confinement-deconfinement transition from the information theoretical viewpoints; we discuss the transition directly from the configurations generated by using the Markov-chain Monte-Carlo (MCMC) method. Particularly, we use the transfer mutual information and the Kullback-Leibler divergence~\cite{kullback1951information} as the information measure; for example, see Ref.~\cite{Kashiwa:2017swa} for the application of the Kullback-Leibler divergence to the Polyakov-loop extended Nambu--Jona-Lasinio model~\cite{Fukushima:2003fw}. It is well known that the mutual information should have the peak at the intermediate ordered region~\cite{barnett2013information}: The mutual information becomes small in the highly disordered state because the system elements are almost independent of each other. In comparison, the mutual information also becomes small in the highly ordered state because each element only has small indeterminacy for the system. The highly ordered and disordered states in the parameter region are corresponding to the deconfined and confined phases realized in QCD, ideally. In other words, QCD has the confinement-deconfinement crossover, but we are interested in whether the topological nontrivial structure exists or not in certain parameter spaces;
there is possibility that the system does not show the thermodynamics singularities like as the topological order, but some qualitative differences exist.

There are several discussions on the center clustering structure of QCD and some related theories~\cite{Gattringer:2010ms,Borsanyi:2010cw,Endrodi:2014yaa} which may be related to the confinement-deconfinement transition and topological structure of the system~\cite{hirakida2020persistent}.
The center clusters are classified from the topological structure in the space of the phase of the Polyakov-loop (Polyakov-line). In the complex Polyakov-loop plane, the spatial distribution of the local Polyakov-loop can have the wide spread.
The distribution of the center cluster depends on the temperature and should have important information of the confinement-deconfinement nature.
Since the mutual information is directly related to the information flow between degrees of freedom (spins in the case of the Potts model) on the nearest-neighbor sites, it is natural to expect that the quantity can be responsible to the center clusters, and also it can care the interaction properties between degree of freedoms.

The information theoretical quantities such as the Shannon entropy, the mutual information, the cross entropy and the transfer information are widely used in several fields such as the financial markets~\cite{harre2009phase,deng2014renyi}, the human collective decision-making~\cite{carbone2015model}, the detection of the phase transition~\cite{lau2013information,barnett2013information} and so on.
There is another interesting information theoretical approach is the persistent homology analysis~\cite{edelsbrunner2000topological,zomorodian2005computing,nakamura2015persistent} which has been applied to the effective Polyakov-line model~\cite{hirakida2020persistent} which has close relation with the Potts model and also QCD;
see Ref.~\cite{Fukushima:2020cmk} for recent interesting criteria for the classification of the confinement-deconfinement crossover.
The persistent homology is recently applied to the condensed matter physics and then it has been reported that the hidden order such as spin nematic ordering and spin liquids can be clarified from the persistent homology analysis~\cite{olsthoorn2020finding}.
Therefore, it seems to be interesting that we can utilize the information theoretic quantity to investigate the mysterious confinement-deconfinement nature of QCD. As a first step to attempt the information theoretical approach to QCD, we start from the QCD effective model with heavy quarks: Actually, the 3-dimensional 3-state Potts model with the external field is employed  in this study because the pure $SU(3)$ Yang-Mills theory has the spontaneous $\mathbb{Z}_3$ symmetry breaking and this nature is included in the Potts model; for example, see Ref.~\cite{wu1982potts} for the review of the model and Refs.~\cite{Alford:2001ug,Kim:2005ck,deForcrand:2010he} for the relation with QCD. Then, the heavy quark contribution can be mapped to the external field which is composed of the quark mass ($M$) and the chemical potential ($\mu$).

In this study, we investigate the behavior of the Kullback-Libeler divergence and the transfer mutual information which represent the information flow in the Markov process and nearest neighbor sites within the several situations by varying the coupling constant, the quark mass and the chemical potential. Then, we can investigate some information theoretical aspects of the confinement-deconfinement nature. Actually, the case of vanishing quark contributions, the pure imaginary chemical potential region and also the real chemical potential region are investigated.

This article is organized as follows. In Sec.~\ref{Sec:TMI}, we explain the formulation of the transfer mutual information and the Kullback-Leibler divergence. The 3-dimensional 3-state Potts model and its setup are explained in Sec.~\ref{Sec:Potts}. Numerical results are shown in Sec.~\ref{Sec:Numerical}. Section \ref{Sec:Summary} is devoted to summary.

\section{Transfer mutual information and Kullback-Leibler divergence}
\label{Sec:TMI}

In this section, we explain some entropic quantities proposed in the (classical) information theory. Also, we propose a new quantity which is so called the transfer mutual information.
The hidden central theme of this section is how we can evaluate those quantities within the Monte-Carlo method.

\subsection{Mutual information}
The mutual information is defined as
\begin{align}
I(A;B) := H(A) - H(A|B),
\end{align}
where $H(X)$ is the Shannon (information) entropy and $H(X|Y)$ means the conditional entropy with random variables, $X$ and $Y$.
The entropies are constructed by using the probability distributions, $0 \le p({\bf s}) \le 1$ with sites $0\le {\bf s} \in \mathbb{Z}$. For example, in the Ising model, each site can have the two degree of freedom (up and down). By using the joint entropy, $H(AB)$, the mutual information is rewritten as
\begin{align}
 I(A;B) = H(A) + H(B) - H(AB),
 \label{MI}
\end{align}
and then the $H(AB)$ is constructed by using the joint probability distribution, $0 \le p({\bf s}_A,{\bf s}_B) \le 1$.
It should be noted that it is difficult to prepare the probability distribution because we need the Monte-Carlo method to perform the integration in the complicated theory, particularly in the quantum field theory.

\subsection{Markov process}
Unfortunately, we cannot easily prepare the probability distribution itself by using the Monte-Carlo method because we cannot calculate the partition function itself. In the case of the 2-dimensional kinetic Ising model, we can access the analytic results of the mutual information, but not in the present model. However, we can prepare the probability distribution (ratio) by considering the stochastic process, $X(t)$ and $Y(t)$ with the fictitious time $0 \le t \in \mathbb{Z}$.
When the probability distribution obeys the Boltzmann distribution and the detailed valance at stationary is manifested, we have
\begin{align}
 P_{t-1 \to t} &= p({s}_i^{(t)}|{s}_i^{(t-1)})
                = \frac{p({s}_i^{(t)},{s}_i^{(t-1)})}{p({s}_i^{(t-1)})},
\end{align}
where $s^{(t)}_i$ is the randomly selected spin at site $i$ at the $t$-th step. In the Markov-chain Monte-Carlo (MCMC) method in the spin model, the random spin flip on randomly chosen site is accepted or rejected by using the probability; see Ref.~\cite{metropolis1953equation}.
In the Metropolis method and the Glauber single spin flip dynamics, the probabilities are given by
\begin{align}
 P_{t-1 \to t}^\mathrm{Metropolis}  &= \mathrm{min} \Bigl(1,e^{-\beta \Delta E} \Bigr), \nonumber\\
 P_{t-1 \to t}^\mathrm{Glauber}  &= \frac{1}{1 + \exp(\beta \Delta E)},
 \label{MCMC}
\end{align}
where $\Delta E$ is the energy difference between the energy at $t$ and
that at $t-1$, and $\beta = 1/T$.
In this study, we employ the Metropolis probability distribution to compute the transfer mutual information.

\subsection{Transfer mutual information}

The transfer entropy operator
\cite{schreiber2000measuring,lau2013information,barnett2013information,deng2014renyi}
is defined as
\begin{align}
 {\hat {\cal T}} &:= -\sum_{s_i,s_j} p(s_i^{(t)}|s_i^{(k)},s_j^{(l)})
                      \ln \frac{p(s_i^{(t)}|s_i^{(k)},s_j^{(l)})}
                      {p(s_i^{(t)}|s_i^{(k)})},
\end{align}
where $t,k,l \in \mathbb{Z}$ indicate the time step of the MCMC process.
Usually, we take $l=t-1$ which is corresponding to the time at the
configuration generation and $k=t-1$.
Also, $i$ and $j$ are taken so as to the nearest
neighbors.
With these settings, $\hat{{\cal T}}$ is the pairwise mutual information
operator and it is denoted as $\hat{{\cal
T}}_\mathrm{pw}$ below.

In this study, we cannot use any analytic expressions of the probability distributions and thus we deform it calculable within the MCMC method as
\begin{align}
 {\hat {\cal T}}_\mathrm{pw}
 &:= - \sum_{s_i} p(s_i^{(t)}|{\bf s}^{(t-1)})
                \ln p(s_i^{(t)}|{\bf s}^{(t-1)})
 \nonumber\\
 &- \sum_{s_j} p(s_j^{(t)}|{\bf s}^{(t-1)})
                \ln p(s_j^{(t)}|{\bf s}^{(t-1)})
 \nonumber\\
 &+ \sum_{s_i,s_j} p(s_i^{(t)},s_j^{(t)}|{\bf s}^{(t-1)})
 \ln p(s_i^{(t)},s_j^{(t)}|{\bf s}^{(t-1)}),
 \label{Eq:TMI}
\end{align}
where $p(s_i^{(t)}|{\bf s}^{(t-1)})$ means the probability that the spin at site $s_i$ is flipped and
$ p(s_i^{(t)},s_j^{(t)}|{\bf s}^{(t-1)})$ does the probability that the spin at site $s_i$ and the nearest neighbor site
$s_j$ are flipped.
The subscript $t-1$ means the configuration generated (fictitious) time.
This quantity can be calculated by using the MCMC method without the analytic expression of the probability distribution.
The first, second and third terms in Eq.~(\ref{Eq:TMI}) are corresponding to $H(A)$, $H(B)$ and $H(AB)$ in Eq.~(\ref{MI}) with the probability distribution in Eq.~(\ref{MCMC}).
Thus, we call it as the {\it transfer mutual information}.
When we evaluate the expectation value of ${\hat {\cal T}}_\mathrm{pw}$, we should take into account all possible pairs of lattice nearest neighbors as
\begin{align}
 {\cal T}_\mathrm{pw}
 &= \frac{1}{3 V} \sum_{\langle i j \rangle}
    \langle {\hat {\cal T}}_\mathrm{pw} \rangle,
\end{align}
where $V$ is the spatial volume of the system and $3V$ is the number of the independent pair of nearest neighbors sites; the lattice spacing is set to $1$ below.
The spin flip is imposed to each configuration after the configuration generation process.
Of course, the transfer mutual information uses the MCMC probability distribution and thus it may depend on the scheme, quantitatively.
It should be noted that the transfer mutual information contains the information flow of the Markov process and also that from the nearest neighbor sites.
In the two-dimensional kinetic Ising model, it is known that the mutual information and the transfer entropy share almost similar properties~\cite{barnett2013information} and thus the present transfer mutual information can be considered as the acceptable quantity for our purpose.
From the mutual information with the above setting, we can treat not only the global structure of the system, but also the local spatial structure of the degree of freedom which is the spin in the Potts model.
Therefore, it is expected that the mutual information is a good quantity to use in the system which has the nontrivial spatial structure.

There is the possibility that we can use some more different probabilities.
Below, we consider QCD as an example.
One choice is using the multiplicity distribution defined as
\begin{align}
g_n &= {\cal Z}_n \xi^n,
\end{align}
where ${\cal Z}_n$ means the canonical partition function with the quark number $n$ and $\xi$ stands for the fugacity.
This quantity can bridge the experimental data and the numerical simulation even in QCD~\cite{Nakamura:2013ska} and it is nothing but the probability distribution of the net quark number.
To make it as the probability distribution, $p_n$, we should consider
\begin{align}
p_n &= \frac{g_n}{\cal N},
\label{Eq:pd_z}
\end{align}
where ${\cal N}$ is  the normalization factor which is corresponding to the grand-canonical partition function itself.
With the probability, there is no scheme dependence, but it needs heavy numerical cost to prepare the canonical partition function which requires the accurate Fourier transformation; see Ref.~\cite{Fukuda:2015mva} as an example.
In addition, by using the probability distribution (\ref{Eq:pd_z}), we can evaluate the information flow between each canonical sector, but does not that between the degree of freedoms.
Therefore, we employ the probability distribution (\ref{MCMC}) appeared in the MCMC process in this study.

\subsection{Kullback-Leibler divergence}

In the information theory, we have another entropy which is so called the relative entropy; a part of the relative entropy is so called the Kullback-Leibler divergence and the operator in the present situation with the Markov process can be expressed as
\begin{align}
 \hat{D}_\mathrm{KL}
 &= \sum_{i} p(s_i^{(t)}|{\bf s}^{(t-1)})
    \ln p(s_i^{(t)}|{\bf s}^{(t-1)}).
\end{align}
It should be noted that this expression seems different from the standard Kullback-Leibler divergence, but the present $p(s_i^{(t)}|{\bf s}^{(t-1)})$ is the conditional probability in the operator and thus is should be suitable to consider the following expectation values with the Boltzmann weight $p(s_i^{(t-1)})$.
The expectation value of the Kullback-Leibler divergence operator is 
\begin{align}
 D_\mathrm{KL} &= \frac{1}{3V}
                  \langle {\hat D}_\mathrm{KL}
                  \rangle
                  \nonumber\\
               &= \frac{1}{3V} \sum_{\bf s} \sum_i p(s_i^{(t)}) \ln p(s_i^{(t)}|{\bf s}^{(t-1)}),
\end{align}
and it represents how similar the probability distributions are.
Thus, the transfer mutual information and the Kullback-Leibler divergence show different aspects of the information about the system.
It should be noted that we replace the probability distributions in $D_\mathrm{KL}$ with the conditional probabilities, we can obtain the mutual information.

For the reader's convenience, we here summarize the properties of the Kullback-Leibler divergence.
The Kullback-Leibler divergence is not bounded above; if the two probability distributions are not overlapped completely ($\forall x,~ p_1(x) \neq p_2(x)$), the Kullback-Leibler divergence becomes $\infty$.
On the other hand, the Kullback-Leibler divergence becomes $0$ if the two probability distributions are the same.
To make the quantity finite in the case with $p_1 (x) \neq p_2 (x) $, we may use the Jensen-Shannon divergence because it does not diverge, $0 \le D_\mathrm{JS} \le \ln 2$~\cite{lin1991divergence}.

It should be noted that the Kullback-Leibler divergence is related to the Fischer's information matrix; with the Taylor expansion, the Kullback-Leibler divergence can be expressed as
\begin{align}
    D_\mathrm{KL} = \frac{1}{2} \sum_{s}\sum_{t} I_{st} \, \delta \lambda_s \delta \lambda_t,
\end{align}
where $\lambda_s,\lambda_t$ mean continuous parameters such as $T$ and $\mu$, and $I_{st}$ is so called the Fischer's information matrix which is the second-order derivative of $D_\mathrm{KL}$ by $\lambda_s$ and $\lambda_t$.
When we replace $D_\mathrm{KL}$ by the free energy, it is corresponding to the thermodynamic geometry recently discussed in QCD~\cite{Castorina:2019jzw,Zhang:2019neb}.
Since the Kullback-Leibler divergence with the MCMC probability should have the thermodynamic information of the theory because the probability distribution is directly related to the thermodynamics, it is interesting to investigate the Kullback-Leibler divergence at finite $T$ and $\mu$.
Actually, the deep relation between the Kullback-Leibler divergence and the phase transition is known fact in statistical mechanics.

\section{QCD effective model with heavy quarks}
\label{Sec:Potts}

In this study, we employ the 3-dimensional 3-state Potts model as the QCD effective model with the heavy quarks~\cite{Alford:2001ug,Kim:2005ck,deForcrand:2010he}.
The Hamiltonian is
\begin{align}
 H &= - \kappa \sum_{{\bf x},{\bf i}}
        \delta_{\Phi_{\mathbf x},\Phi_{{\bf x+i}}}
      + \sum_{\bf x}
        \Bigl ( h_+ \Phi_{\bf x} + h_- {\bar \Phi}_{\bf x} \Bigr),
\end{align}
where ${\bf i}$ means the unit vector in the three dimensional space, $\kappa$ means the coupling constant, $h_{\pm}$ denotes the external field and $\Phi_{\bf x}$ (${\bar \Phi_{\bf x}}$) is the ${\mathbb Z}_3$ values (its conjugate) on each site; it is corresponding to the Polyakov-loop in QCD.
The external fields are expressed with the quark mass and the chemical potential as
\begin{align}
h_\pm &= e^{-\beta(M\mp\mu)},
\end{align}
which is induced from the fermion determinant in QCD partition function; see appendix\,\ref{Sec:appendix} for details.
If the quark mass is sufficiently heavy in QCD, the 3-dimensional 3-state Potts model with the external field can be treated as the effective model of QCD.
Recently, the extension of the Potts model has been done by considering the $\mathbb{Z}_3$ symmetrization and used to investigate the sign problem~\cite{Hirakida:2016rqd}; see Ref.\,\cite{deForcrand:2010ys} for details of the sign problem.
This means that the analysis of the Potts model is still important to understand some QCD properties.

At nonzero real $\mu$, the Hamiltonian of the 3-state Potts model is no longer real at each MCMC process and then we cannot consider that $e^{-\beta H} $ is probability distribution which leads the Boltzmann distribution at stationary.
One choice to overcome the difficulty is using the phase quenched probability distribution in the configuration averaging procedure and employing the reweighting.
However, if the average phase factor becomes smaller, it is very difficult to obtain reliable results.
In this paper, thus, we consider the small real chemical potential because the sign problem is not serious.
Also, we introduce the imaginary chemical potential because the Hamiltonian goes back to real values.

To calculate ${\cal T}_\mathrm{pw}$ and $D_\mathrm{KL}$, we utilize the MCMC method.
We here use the Metropolis method to perform the spin flip process and then we generate $10^4$ configurations analyzed each $100$ updation after the thermalization.
Simulations are performed with $V=6^3$, $8^3$ and $10^3$.
Errors are estimated by using the Jack-Knife method with the bin size $N_\mathrm{bin}=10$.

\section{Numerical results}
\label{Sec:Numerical}

We show our numerical results with the real and imaginary external fields and without the external field by using the Monte-Carlo method.

\subsection{Vanishing $h_\pm$}

In this situation, we have the exact $\mathbb{Z}_3$ symmetry in the
Hamiltonian level and then we always have $\langle \Phi \rangle=0$ in principle; even if the symmetry spontaneously broken, the configuration averaging procedure leads $\langle \Phi \rangle=0$.
It should be noted that we can take the absolute value of the spacial averaged $\Phi$ before taking the configuration average and then the expectation value becomes still nonzero;
we abbreviate it as $|\Phi|$ below.
In this study, we introduce the extremely small external field to make the spin has a unique direction after the spontaneous symmetry breaking; it is the standard procedure to detect the phase transition.
Also, we do not take extrapolation to the thermodynamic limit and thus the Polyakov-loop still has the non-zero value in the confined phase; it is nothing but the finite size artifact.
\begin{figure}[b]
 \centering
 \includegraphics[width=0.22\textwidth]{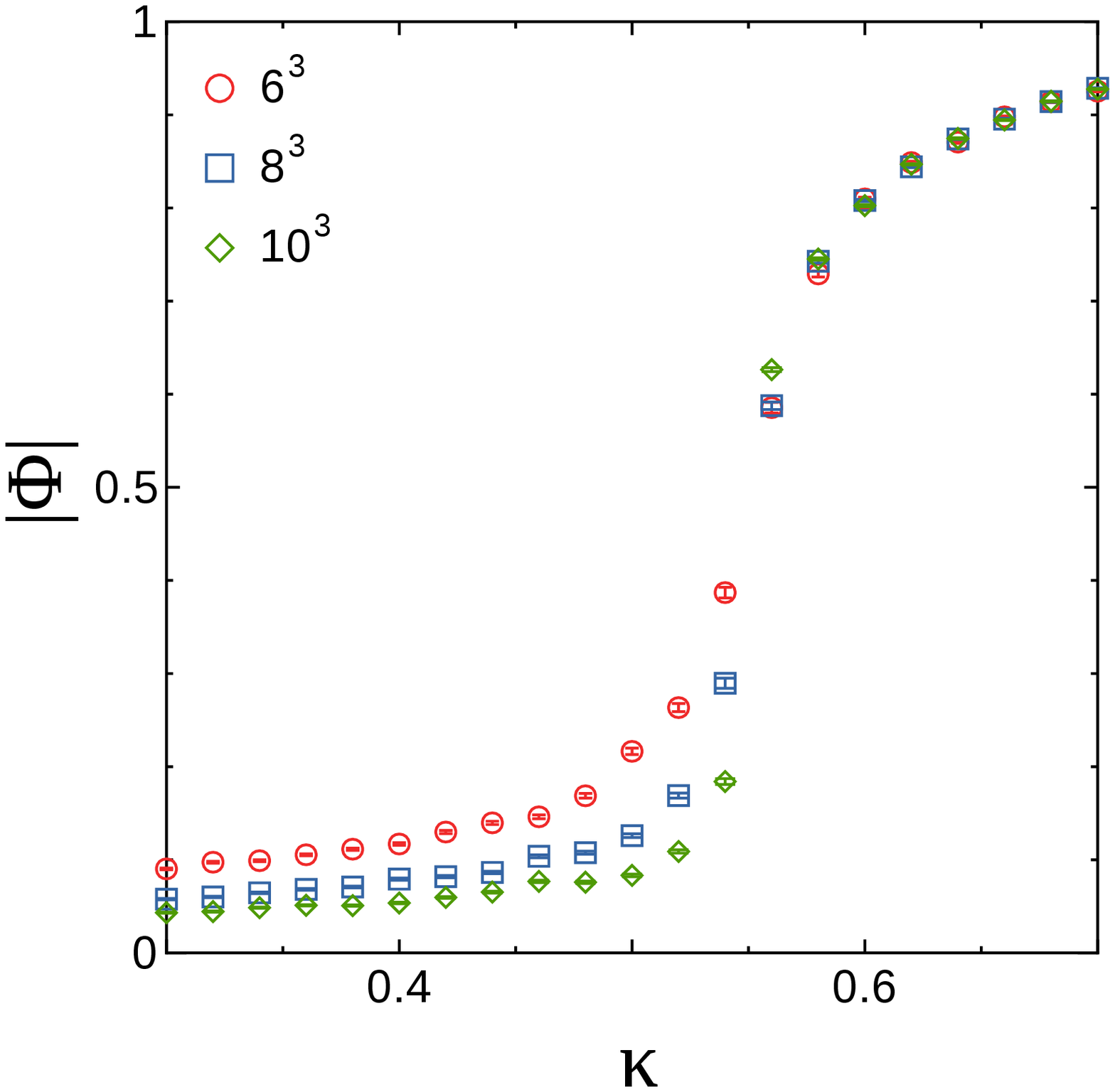}
 ~~
 \includegraphics[width=0.225\textwidth]{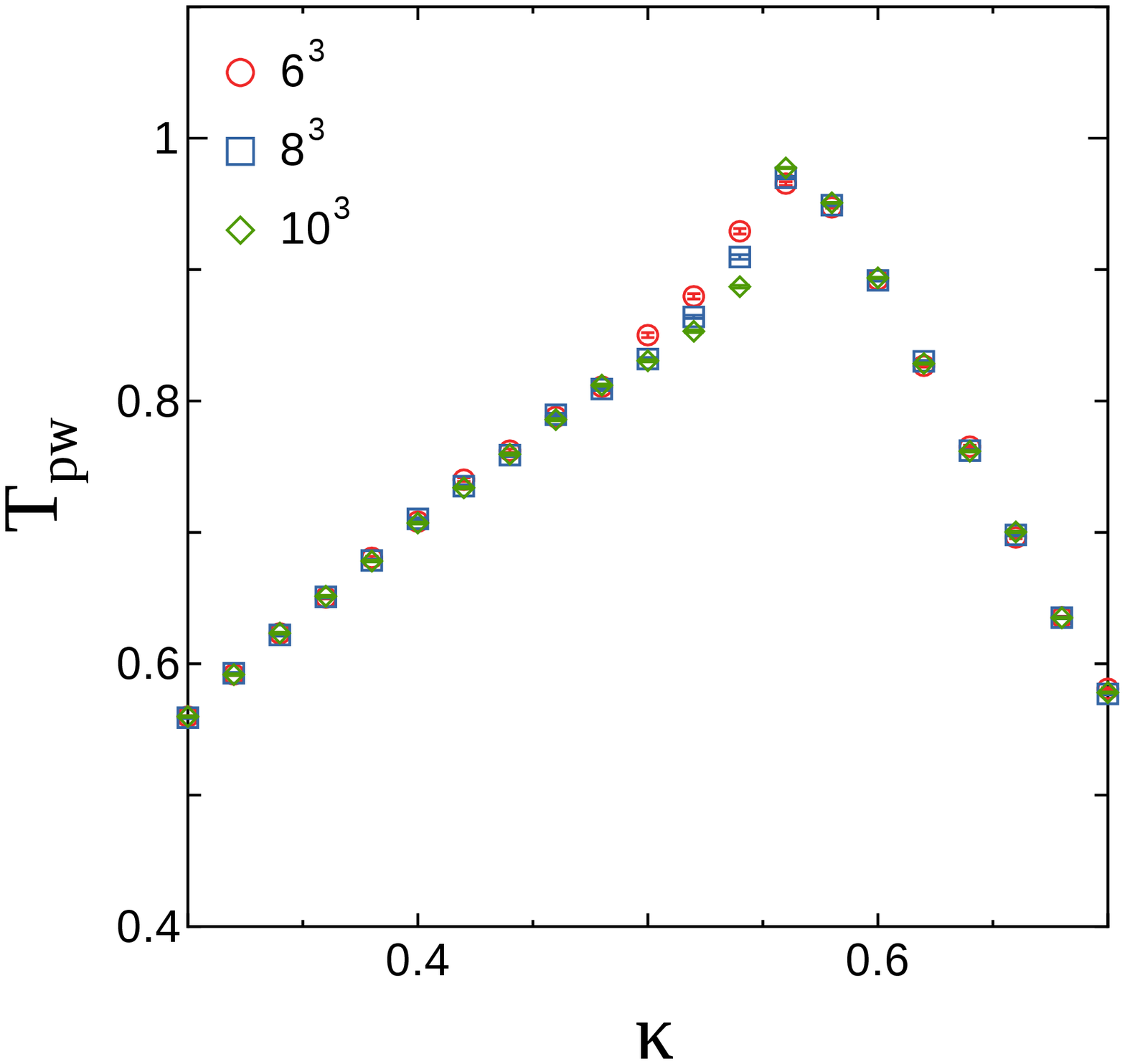}
 \nonumber\\
 \hspace{10mm}
 \\
 \includegraphics[width=0.22\textwidth]{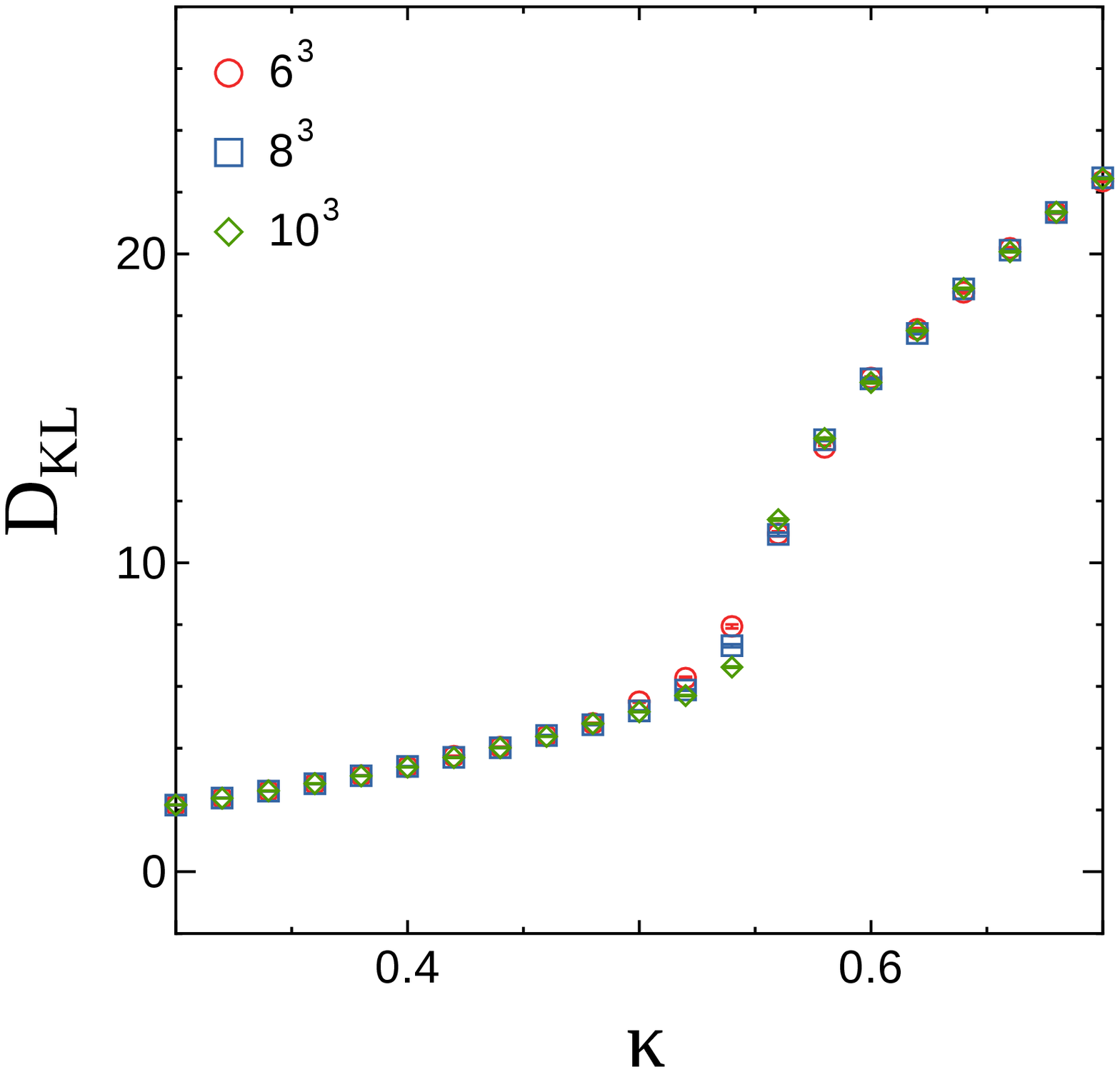}
 \caption{
 The $\kappa$-dependence of the
 $|\Phi|$, ${\cal T}_\mathrm{pw}$ and $D_\mathrm{KL}$ at $\beta=1$
 with $V=6^3$, $8^3$ and $10^3$.}
\label{Fig:Vol}
\end{figure}

Figure~\ref{Fig:Vol} shows the $\kappa$-dependence of
$|\Phi|$, ${\cal T}_\mathrm{pw}$ and $D_\mathrm{KL}$ at $\beta=1$.
The peak exists in the ${\cal T}_\mathrm{pw}$ and it appears very close
tho the rapidly changing point of $|\Phi|$ and $D_\mathrm{KL}$.
This indicates that the transfer mutual information picks up the
fluctuation of the system from the spin configuration space.
Also, the finite size effects in the transfer mutual information seems to
be smaller than that in the Polyakov-loop.
Since the present mutual information is consist of the information flow between nearest neighbor sites and thus the finite size effect is expected to be small.

\subsection{Nonzero real $\mu$}
\label{Sec:hh}

In this case, the 3-dimensional 3-state Potts model has the sign problem and it becomes more serious when the chemical potential becomes larger. To circumvent the problem, we here use the reweighting method~\cite{Fodor:2001au,Fodor:2002km,Fodor:2001pe,Fodor:2004nz}; we make the probability distribution by using the phase quenched one and the observable quantities are reweighted to make it correctly.
Actually, we prepare the MCMC probability by using the absolute value of the original Boltzmann weight, $|e^{-\beta H}|$, with the average phase factor, $e^{-\beta H}/|e^{-\beta H}|$.
However, when the average phase factor which represents the difference between the original and phase quenched probability distributions becomes smaller, we cannot obtain reliable results and thus we can not consider whole chemical potential region. We here consider $\mu < M$.
In the Potts model, there are several modern methods to circumvent the sign problem, but we are interested in the information flow and thus we still use the standard Metropolis algorithm; for example, see Ref.\,\cite{deForcrand:2017rfp} for smart approaches to the sign problem in the Potts model.

\begin{figure}[t]
 \centering
 \includegraphics[width=0.22\textwidth]{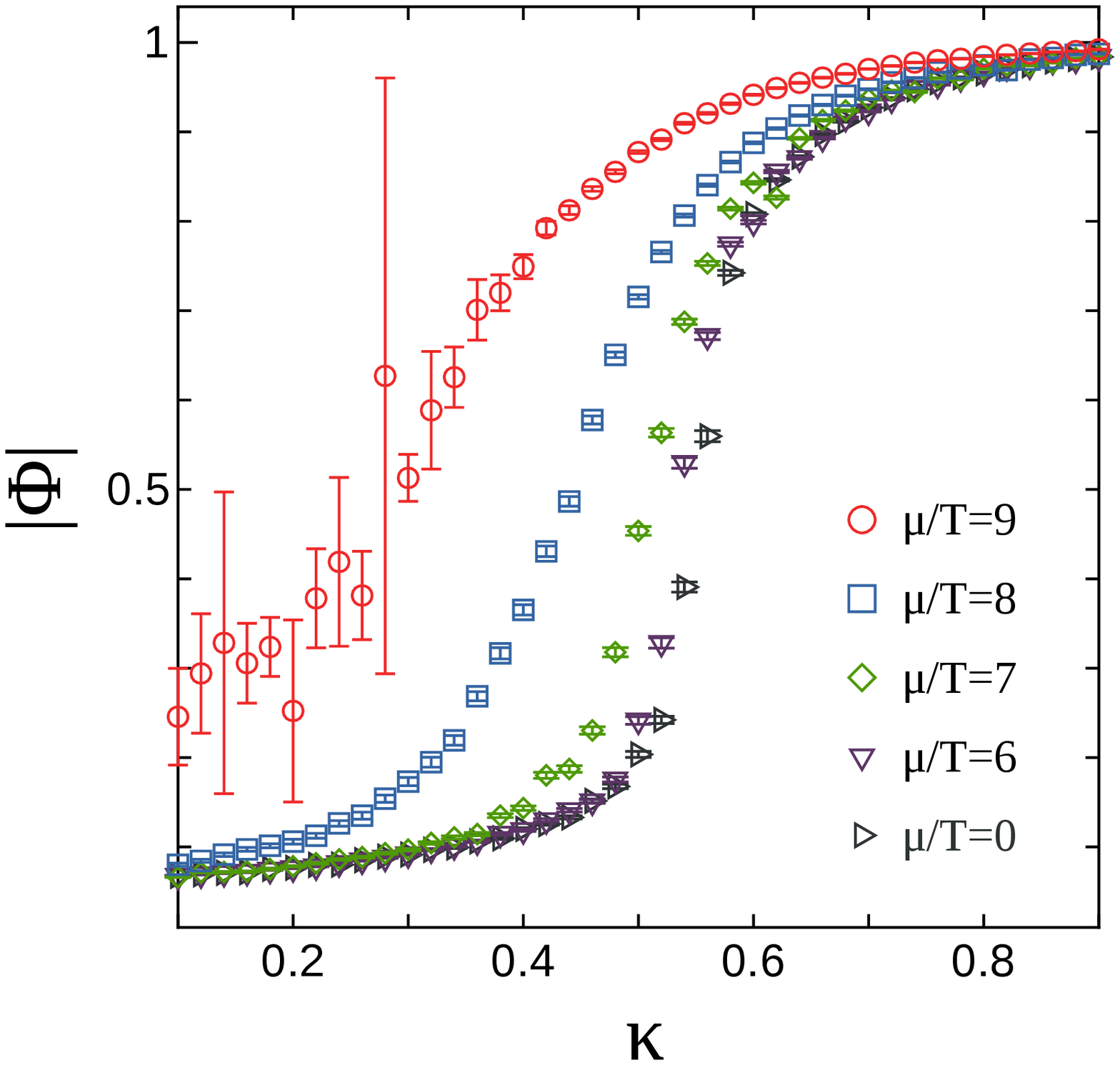}
 ~~
 \includegraphics[width=0.22\textwidth]{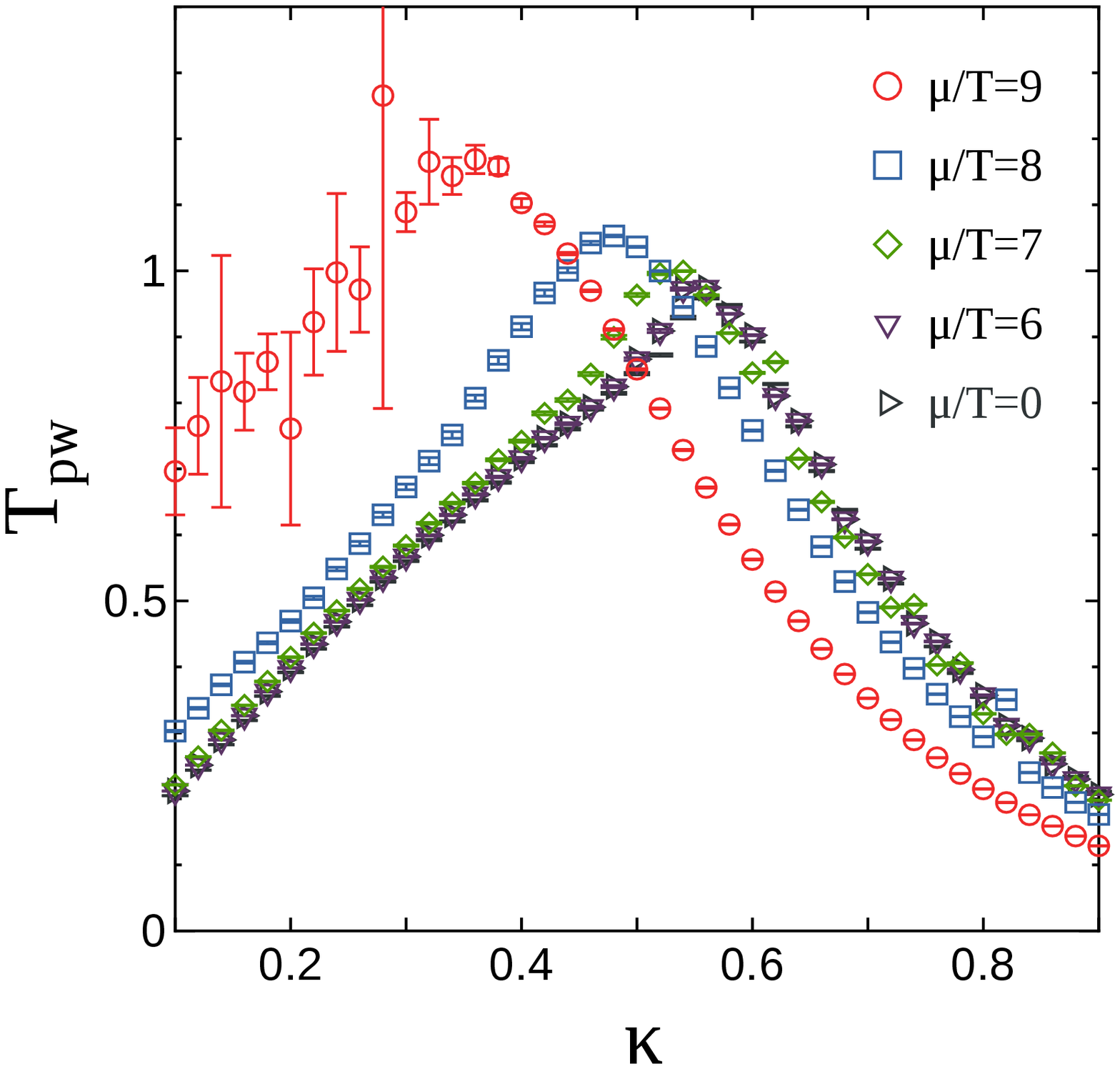}
 \nonumber\\
 \hspace{10mm}
 \\
 \includegraphics[width=0.22\textwidth]{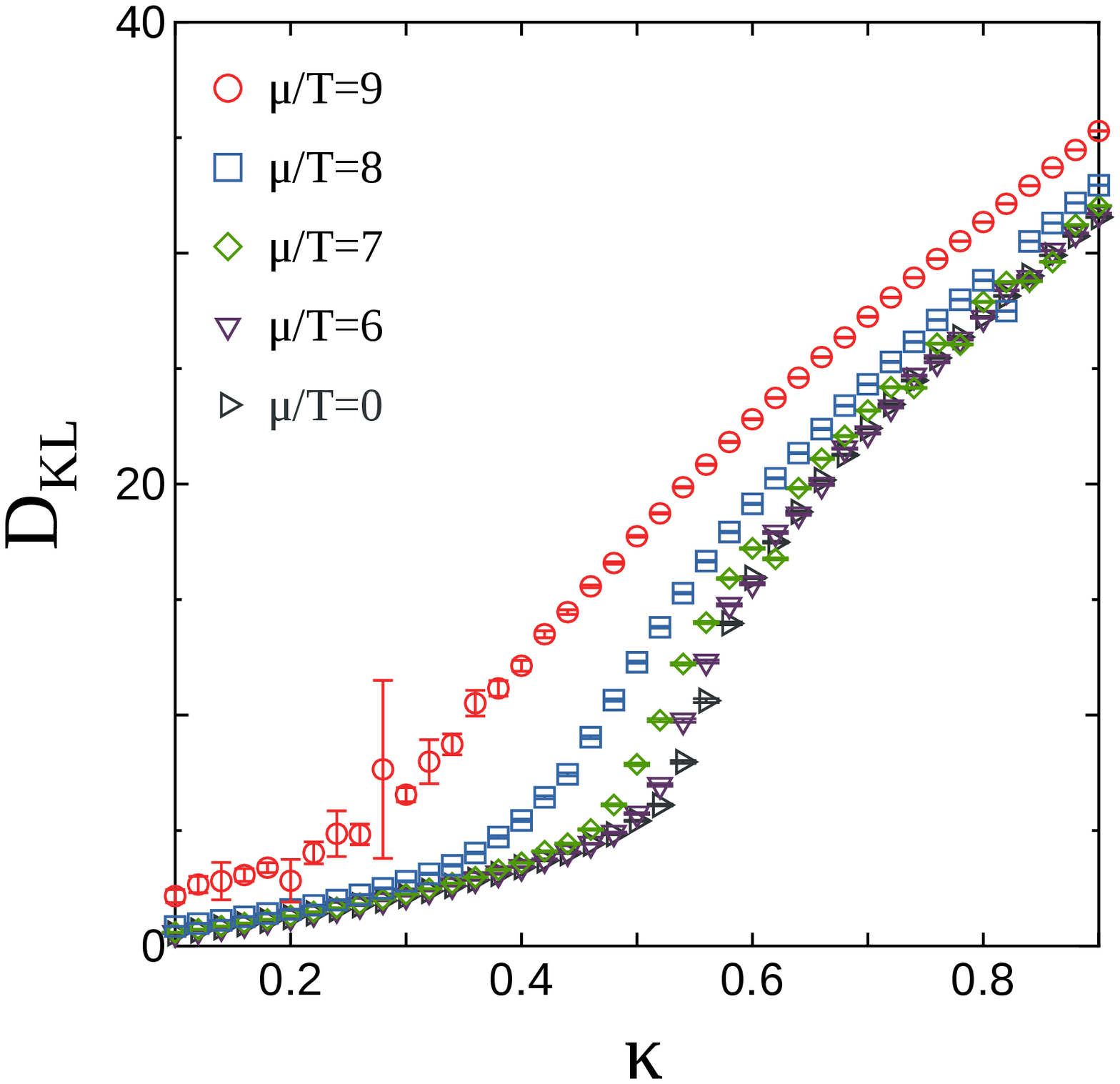}
 \caption{
 The $\kappa$-dependence of $|\Phi|$, ${\cal T}_\mathrm{pw}$ and
 $D_\mathrm{KL}$ where $\beta=1$ with $M=10$ and $V=6^3$, respectively.
 }
\label{Fig:Re}
\end{figure}
Figure~\ref{Fig:Re} shows the $\kappa$-dependence of $|\Phi|$, ${\cal T}_\mathrm{pw}$ and  $D_\mathrm{KL}$ with $\beta=1$, $M=10$ and $V=6^3$.
Because of the small average phase factor at finite $\mu$, the error bar becomes large, but we can see that the rapidly changing point of $|\Phi|$ decreases with increasing $\mu$.
The smoothing behavior of the Polyakov loop with increasing $\mu/T$ is the opposite behavior comparing with the chiral transition in QCD.
The Kullback-Leibler divergence shows similar behavior with the Polyakov-loop.
Interestingly, the peak position of ${\cal T}_\mathrm{pw}$ shifted to lower $\kappa$ with increasing $\mu$ and the peak strength becomes strong.
This strongly indicates that the correlations between nearest neighbor sites is enhanced.
Also, the transfer mutual information feels effects of $\mu$ in the confined phase rather than the deconfined phase.
This means that the confinement phase has strong information flow coming from fermion effects.

\subsection{Nonzero pure imaginary $\mu$ case}
\begin{figure}[t]
 \centering
 \includegraphics[width=0.22\textwidth]{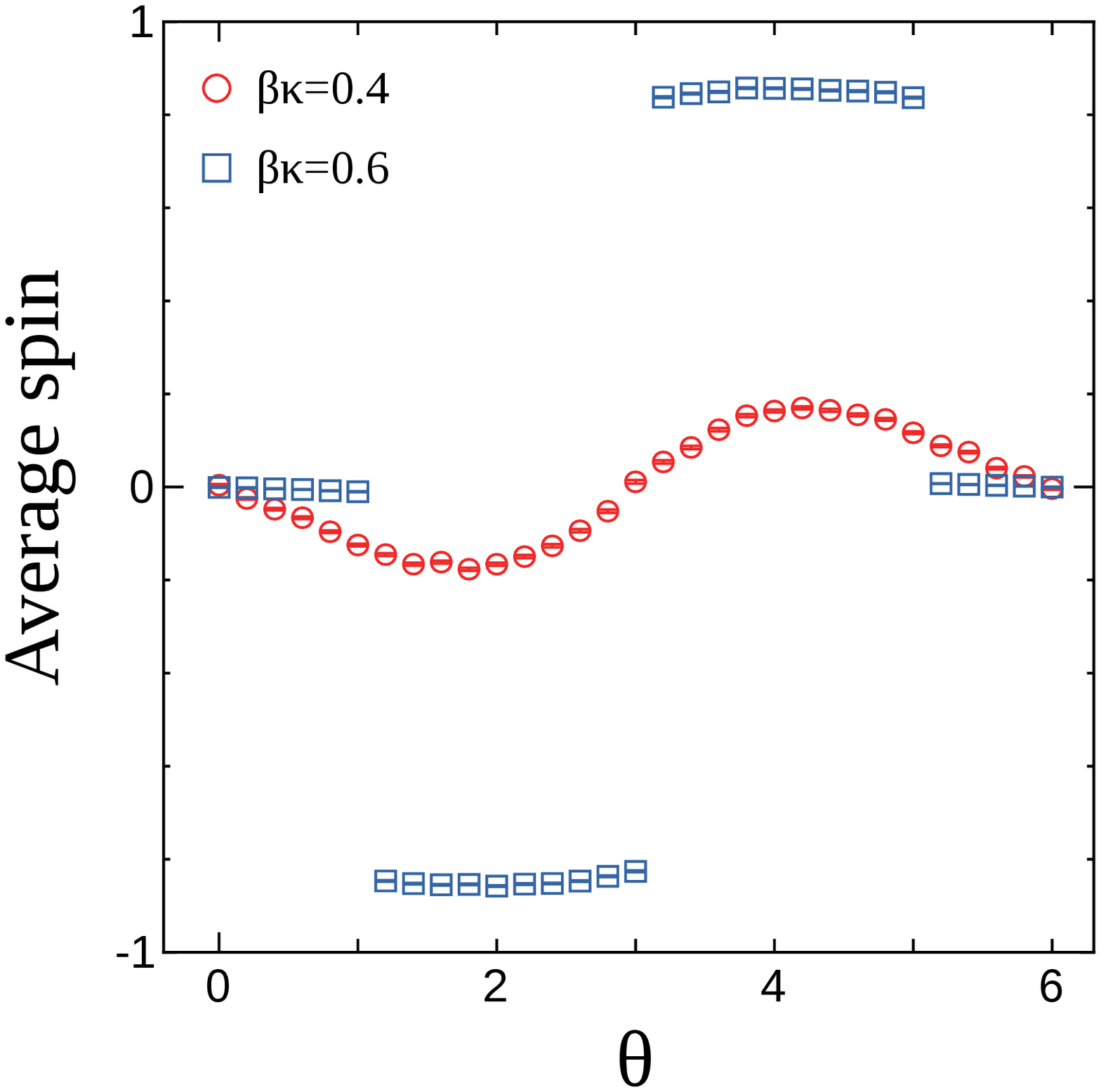}
 ~~
 \includegraphics[width=0.23\textwidth]{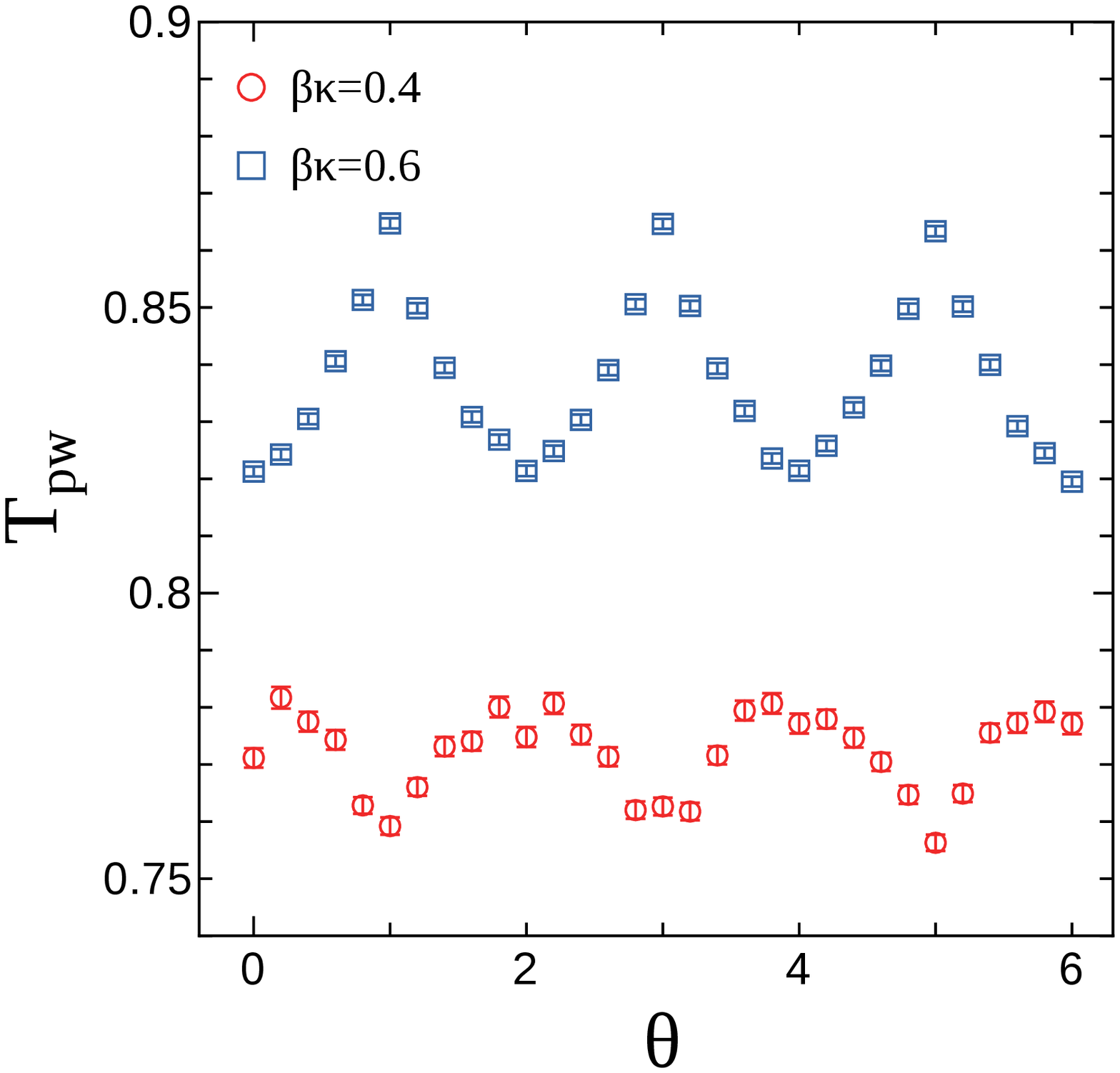}
 \nonumber\\
 \hspace{10mm}
 \\
 \includegraphics[width=0.23\textwidth]{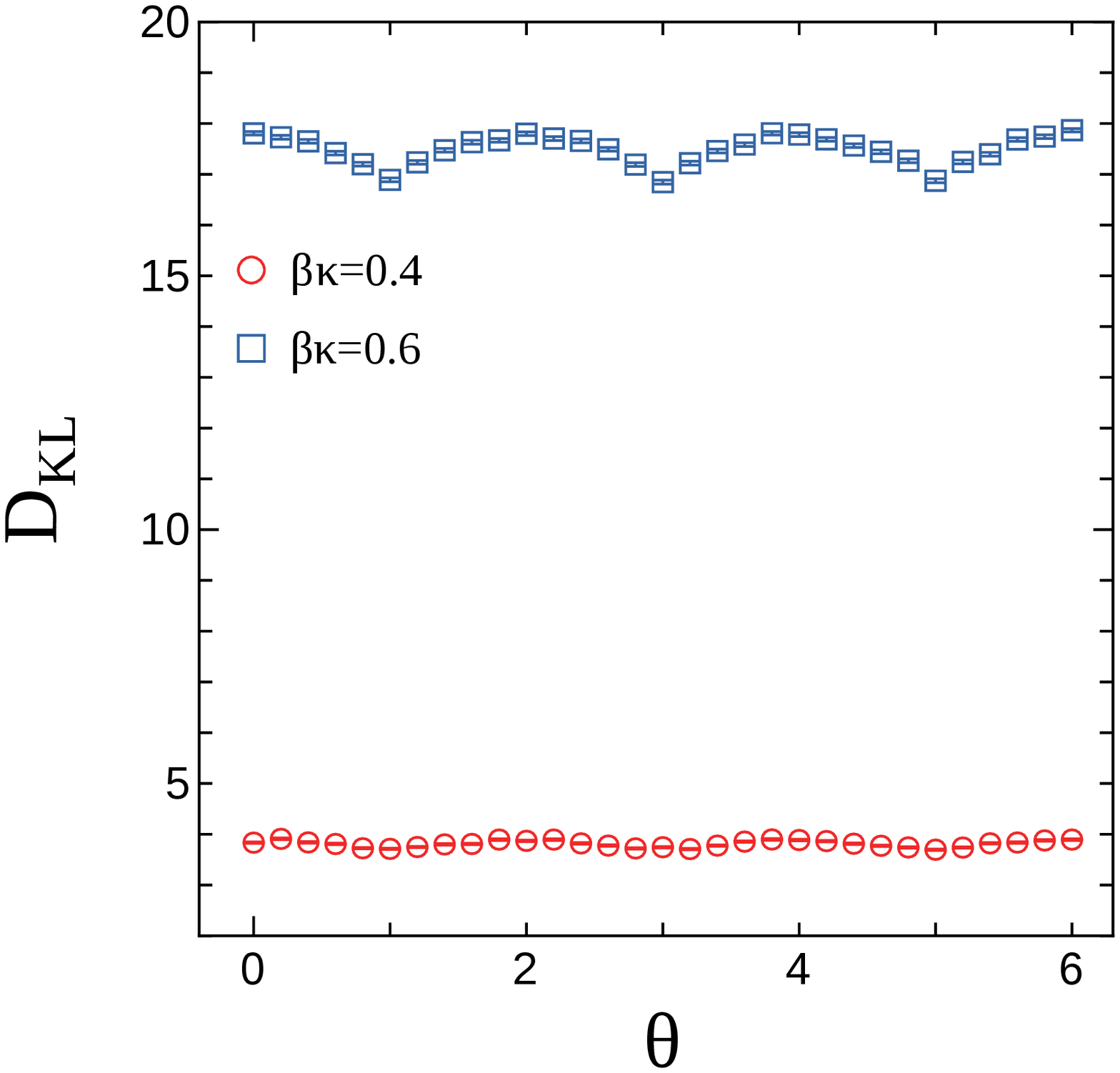}
 \caption{
 The $\theta$-dependence of the average spin, ${\cal T}_\mathrm{pw}$ and
 $D_\mathrm{KL}$ with $V=6^3$.
 The circle and square symbols are results with $\beta \kappa=0.4$ and $0.6$ with $\beta=0.1$ and $M=10$,
 respectively.
 }
\label{Fig:Im-as}
\end{figure}

Finally, we consider the pure imaginary chemical potential.
The imaginary chemical potential can be interpreted as the flux insertion to the fictitious hole of the time direction in the QCD with the imaginary time formulation and it has the analogy with the way used in the condensed matter physics to calculate the Berry phase.
In this case, we do not encounter the sign problem and also there is the Roberge-Weiss transition along the $\theta:=\mu/(iT)$ direction in addition to the confinement-deconfinement transition~\cite{Roberge:1986mm,D'Elia:2002gd,deForcrand:2002ci,deForcrand:2003hx,D'Elia:2004at}.
Also, there may be the possibility that we can pick up some information of the confinement-deconfinement transition from this region~\cite{Kashiwa:2015tna,Kashiwa:2016vrl,Kashiwa:2017yvy}.
It should be noted that the following results are first result which explicitly shows the detailed behavior of the Roberge-Weiss periodicity and the transition in the Potts model with the external field.
The $\theta$-dependence of the average spin with $\beta \kappa=0.4$ and $0.6$ with $\beta=0.1$ is shown in the top-left panel of Fig.~\ref{Fig:Im-as}.
To clearly see the oscillating behavior, we here take $\beta=0.1$ and $M=10$.

In the Potts model, spin itself is more fundamental quantity comparing with the Polyakov-loop and thus we here calculate
\begin{align}
 \langle \mathrm{Average~spin} \rangle
 &= \Bigl\langle \frac{1}{V}\sum_{\mathbf{x}} \mathrm{spin}(\mathbf{x}) \Bigr\rangle,
\end{align}
where we assign spins as $0$, $+1$ and $-1$ for $\mathrm{arg}(\Phi)=0$,
$\mathrm{arg}(\Phi)=2\pi/3$ and $\mathrm{arg}(\Phi)=4\pi/3$, respectively.
We can clearly see that there is the RW transition, if $\kappa$ is sufficiently large where $|\Phi|$ has large value.
The top-right and bottom panels of Fig.~\ref{Fig:Im-as} shows the $\theta$-dependence of ${\cal T}_\mathrm{pw}$ and $D_\mathrm{KL}$, respectively.
We can see that the transfer mutual information and the Kullback-Leibler divergence show the RW periodicity and the tendency of the RW transition when the $\mathbb{Z}_3$ symmetry is strongly broken.
Interestingly, the oscillation of the transfer mutual information becomes significantly different behavior below and above RW endpoint.
This indicates that information flow is different and it has the additional information about the confinement-deconfinement nature comparing with the Polyakov-loop and the Kullback-Leibler divergence.

\section{Summary}
\label{Sec:Summary}

In this study, we have investigated the information theoretical aspects of the confinement-deconfinement nature in the QCD effective model with heavy quarks; the 3-dimensional 3-state Potts model is employed.
We have considered the transfer mutual information which represents the information flow in the Markov process and the lattice nearest neighbor sites in the case of the pure gauge limit, the finite imaginary chemical potential region and the finite real chemical potential region to understand the confinement-deconfinement transition.
The transfer mutual information can pick up the confinement-deconfinement transition of the system and then it has the peak around  the transition point unlike the Kullback-Leibler divergence which has the rapidly changing point around the point in the finite size system. Also, it can clarify the Roberge-Weiss transition and its endpoint which is important to understand QCD properties at finite density.

At finite real chemical potential, we employ the reweighting method and calculate the Polyakov-loop, the transfer mutual information and the Kullback-Leibler divergence.
We have found that the value of $\kappa$ at the peak position of the transfer mutual information decreases with increasing the real chemical potential.
Also, the transfer mutual information still has the peak position if the system indicates the crossover from the Polyakov-loop behavior.
This means that there is the region that the information transfer still active even when the Polyakov-loop indicates the crossover behavior of the thermal system.
It may mean that there is the possibility that the significant system changes which are not characterized by the order parameter of the spontaneous symmetry breaking exist.
To clarify it, we need more studies on the confinement-deconfinement transition from the information theoretical approach, topological approach and the ordinary approach.
Particularly the Uhlmann phase~\cite{Uhlmann1986,Viyuela2014} may be the promising quantity to investigate the confinement-deconfinement transition since it is the extended quantity of the Berry phase to the quantum mixed-state~\cite{Uhlmann1986} and it can describe the finite temperature topological order~\cite{Viyuela2014}.
These will be discussed in the future work~\cite{KD}.

When we apply the transfer mutual information to QCD, we need some more extension since the degree of freedom in QCD is not the discrete spin but the $SU(3)$ gauge field.
However, the field still has the $\mathrm{Z}_3$ symmetry in the classical action level in the pure gauge limit and thus we can define the $\mathbb{Z}_3$ spin flip which has the perfect analogy with the spin flip in the 3-state Potts model.
Therefore, we can calculate the transfer mutual information in QCD by using the $\mathbb{Z}_3$ flip (transformation) employing the Metropolis update procedure. We will report it elsewhere.
Particularly, QCD has another interesting nature which is so called the chiral symmetry breaking and thus it is interesting how the transfer mutual information feels the phenomena.

 \begin{acknowledgments}
K.K. is grateful to Y. Kikuchi for discussions on the relation between the Potts model and QCD.
This work is supported in part by the Grants-in-Aid for Scientific Research from JSPS (No. 18K03618, 19H01898 and 20K03974).
 \end{acknowledgments}

\appendix
\section{Mapping of $m$ and $\mu$ to external field of Potts model}
\label{Sec:appendix}

Based on Ref.~\cite{McLerran:1981pb}, we can understand how the Potts model has the relation with QCD, particularly how to map the quark mass and the chemical potential in QCD to the external field in the Potts model.

With the sufficiently heavy quark mass, quarks can be treated as the static quark and then quarks obey the time-evolution equation;
\begin{align}
&\Bigl(\frac{\partial}{\partial \tau} + i A_4 + \gamma_0 m \Bigr)
    \, q({\vec x},\tau) =0
\nonumber\\
&\hspace{1cm}\to
 \Bigl[ \Bigl(\frac{\partial}{\partial \tau} + A_4\Bigr)^2
       - m^2 \Bigr]
    \, q({\vec x},\tau) =0,
\end{align}
where we consider the Euclidean space-time.
To introduce $\mu$, we should replace $A_4$ with $A_4 - i \mu$.
The solution of the equation becomes
\begin{align}
 q({\vec x},\tau) &= {\cal T} \exp \Bigl[ i \int_0^\beta A_4 d \tau -\beta (m-\mu) \Bigr]\, q({\vec x},0),
\end{align}
where ${\cal T}$ means the time-ordering operator and we take the solution $\exp(-\beta m)$ instead of $\exp(\beta m)$ to match the pure gauge limit.

The partition function with static $N_q$-quarks and $N_{\bar q}$-antiquarks is
\begin{align}
&\sum_{|s\rangle} \langle s| e^{-\beta {\cal H}} |s\rangle
\nonumber\\
& = \mathrm{tr_c}
    \Bigl[ e^{-\beta {\cal H}}
           L({\vec x}_1) \cdots L({\vec x}_{N_q})
           L^\dag({\vec x}_1) \cdots L^\dag({\vec x}_{N_{\bar q}})
    \Bigr]
\nonumber\\
&\sim {\cal Z},
\end{align}
where $|s\rangle$ means the state with static $N_q$-quarks and $N_{\bar q}$-antiquarks, ${\cal H}$ is the pure gauge Hamiltonian which is modeled by the standard Potts model and $L$ means the modified operator of the Polyakov loop defined as
\begin{align}
    L({\vec x}) &= \frac{1}{N_\mathrm{c}} \mathrm{tr_c}
    {\cal T} e^{ i \int_0^\beta A_4 d\tau} e^{-\beta (m-\mu) },
\end{align}
where $\mathrm{tr_c}$ means the trace acting in the color space.
From the result, we can read off the effective Hamiltonian for the QCD with heavy quarks which is corresponding to the Potts Hamiltonian with the external field; we can understand how the Potts spin correlates with the quark mass and the chemical potential after taking sums of the number of quark and antiquark, $h_\pm = \exp\beta[-(m\mp\mu)]$.
For more details except the present mapping, see Ref.~\cite{Alford:2001ug}.
It should be noted that we can understand why the Polyakov-loop is related to the free energy for the single-quark excitation in the static limit and why the relation cannot be manifested with dynamical quarks from the equations.

\bibliography{ref.bib}

\end{document}